\documentclass[11pt]{article}
\usepackage[mathcal,mathscr]{eucal}
\usepackage{latexsym,amssymb}
\usepackage{amssymb}
\newcommand{\be}{\begin{equation}}
\newcommand{\ee}{\end{equation}}
\newcommand{\ba}{\begin{eqnarray}}
\newcommand{\ea}{\end{eqnarray}}

\renewcommand{\Im}{\mathrm {Im} \,}

\font\Eul = eufm10

\font\Eul = eufm7 at 12pt 

\def\bC{{\mathbb C}}
\def\bR{{\mathbb R}}
\def\bN{{\bf N}}
\def\Im{{\rm Im\,}}

\def\BB{{\cal B}}

\def\DD{{\cal D}}

\def\HH{{\cal H}}

\def\RR{{\cal R}}

\def\TT{{\cal T}}

\def\WW{{\cal W}}

\def\ovl{\overline}

\def\Gd{L_+^\uparrow}
\def\Rep{{\bf \Phi}}

\def\bea{\begin{eqnarray}}
\def\eea{\end{eqnarray}}
\def\le{\left}
\def\ri{\right}

\def\RR{\mathbb R}

\title{\bf The asymptotic symmetry of de Sitter spacetime}
\author{Jacques Bros$^{\rm a}$,
         Henri Epstein$^{\rm b}$,
         Ugo Moschella$^{\rm c}$,\\[3pt]
$^{\rm a}$ {\small Service de Physique Th\'eorique, C.E. Saclay,
91191 Gif-sur-Yvette, France} \\
{$^{\rm b}$ \small Institut des Hautes Etudes Scientifiques, 91440
Bures-sur-Yvette, France} \\
$^{\rm c}$ {\small  Dipartimento di
Scienze Matematiche Fisiche e
                   Chimiche,}\\
       {\small Universit\`a dell'Insubria, 22100 Como and INFN sez. di Milano, Italy} \\
        }
\begin{document}
\maketitle

\abstract We show how to construct a set of Euclidean conformal
correlation functions on the boundary of a de Sitter space from
an interacting bulk quantum field theory with a certain asymptotic
behaviour. We discuss the status of the boundary theory w.r.t. the
reflection positivity and conclude that no obvious physical
holographic interpretation is available.

\newpage

\section{Introduction}

It has been recently proposed a duality between a quantum theory
on de Sitter space and a euclidean theory on its boundary
\cite{stro} which should encode the de Sitterian quantum gravity
degrees of freedom \cite{stro,witten}. In this paper we show that
one can associate with a general (scalar) de Sitter quantum field
theory satisfying suitable condition a conformal euclidean field
theory on the boundary, here identified with a copy of the cone
asymptotic to the de Sitter manifold in the embedding spacetime.
However, the field theory that one gets this way does not in
general satisfy reflection positivity, which is required to admit
a physical interpretation. Therefore, the proposed construction
can have in general a technical interest but no obvious
holographic interpretation seems to be available.

\section{Notations and geometry}
Let us consider  the vector space ${\mathbb R}^{d+1}$ equipped
with the Lorentz scalar product:
\begin{equation}
X\cdot X'  = {X^0} {X'}^0 - {X^1} {X'}^1 - \cdots - {X^d}{X'}^d \
. \label{ambientmetric}
\end{equation}
The $d$-dimensional dS universe can then be identified with the
quadric
\begin{equation}
dS_{d} = \{ X \in \mathbb R^{d},\;\; {X^2}=-R^2\},
\end{equation}
where  $X^2= {  X} \cdot{X}$, endowed with the induced metric
\begin{equation}
{\mathrm d}s^2 = \left.\left(d{{X}^0}^{\,2}-d{{X}^1}^{\,2} -
\cdots - d{{X}^{d}}^{\,2}\right) \right|_{dS_{d}}.
\label{metric}
\end{equation}
The future cone is defined in the real Minkowski space
$\bR^{d+1}$ as the subset $$ V_+ = -V_- = \{X\in \bR^{d+1}\ :\
X^{0} > 0,\ \ X\cdot X >0 \} $$ and the future light cone as $C_+
= \partial V_+ = -C_-$. The future cone induces the (partial)
causal order defined by $\overline{V_+}$, i.e. $X \leq Y $ if and
only if $ Y-X \in \overline{V_+}$ as vectors in the ambient
space. The future and past shadows of a given event $X$ in $dS_d$
are given by $$\Gamma^+{(X)}= \{Y \in dS_d: Y \geq X \},\ \
\Gamma^-{(X)}= \{Y \in dS_d: Y \leq X \}.$$ If $X^2 = -R^2$ and
$\eta^2 = 0$, then $(X+\eta)^2 = -R^2$ is equivalent to $x\cdot
\eta=0$, and remains true if $\eta$ is replaced with $t\,\eta$
for any real $t$. Hence the boundary set
\begin{equation}
{\partial \Gamma}(X) =\{Y\in dS_d: (X-Y)^2=0\}
\end{equation}
of $\Gamma^+{(X)}\cup\Gamma^-{(X)}$ is a cone (``light-cone'')
with apex $X$, the union of all linear generators of $dS_d$
containing the point $X$. Two events $X$ and $Y$ of $dS_d$ are in
space-like separated if $Y \not\in
\Gamma^+{(X)}\cup\Gamma^-{(X)}$, i.e. if $ X\cdot Y > -R^2$.

The symmetry group of the de Sitter space-time, is the connected
Lorentz group of the ambient Minkowski space, i.e.
$\Gd=SO_{0}(1,d)$ leaving invariant  each of the sheets of the
cone $C = C_{+}\cup C_{-}$. $\Gd$ acts transitively on $dS_d$.

We will also consider the complexification of $dS_{d}$:
\begin{equation}
dS^{(c)}_{d} = \{ Z = X+iY \in \mathbb C^{d}, \;\;Z^{2}= -R^2\}.
\end{equation}
In other terms, $Z = X+iY$ belongs to $dS^{(c)}_{d}$ if and only
if $X^2 - Y^2 = R^2$ and $X\cdot Y = 0$. The complex Lorentz group
$L_+(\bC)$ acts transitively on $dS_d^{(c)}$.

The familiar forward and backward tubes are defined in complex
Minkowski space as ${\rm T}_{\pm} = \bR^{d+1} \pm iV_+$, and we
denote their intersection with the de Sitter manifold as follows:
\begin{equation}
\TT_{+} = \makebox{\rm T}_+\cap dS^{(c)}_{d},\;\;\;\;\; \TT_{-} =
\makebox{\rm T}_-\cap dS^{(c)}_{d}. \label{tubi1}
\end{equation}
 Since $\overline{{\rm T}_+}\cup \overline{{\rm T}_-}$ contains
the ``Euclidean subspace'' of the complex Minkowski spacetime
${{\mathbb C}}^{d+1}$, that is  ${\mathbb E}^{d+1}=\{
Z=(iY^{0},X^{1},\ldots, X^{d}): (Y^{0}, X^{1},\ldots, X^{d}) \in
{{\mathbb R}}^{d+1} \}$  the subset $\overline{\TT_+}\cup
\overline{\TT_-}$ of $dS_d^{(c)}$ contains the  ``Euclidean
sphere'' ${ S}_{d}=\{Z=(iY^{0},X^{1},\ldots, X^{d}):\;\;
{Y^{(0)}}^{2}+{X^{(1)}}^{2}+\ldots+{X^{(d)}}^{2}=R^{2}\}$.

The de Sitter manifold admits a global parametrization  $X =
X[\tau, {\omega}]$ whose ``constant time'' sections $S(\tau)$ are
spheres:
\begin{equation}
\left\{
\begin{tabular}{lclc}
$X^{0}$  &=& $ \sinh \tau $ & \cr
$X^{i} $  &=& $\cosh \tau \ {\omega}^i $ & $i=1,...,d$, \ \
${\omega}^2 = {{\omega}^1}^2+ \cdots +{{\omega}^{d}}^2=1$
\end{tabular}\right. .
\label{sphericcoordinates}
\end{equation}
This parametrization has the advantage to globally describe the
real de Sitter manifold. Another useful parametrization is the
{``horocyclic parametrization'' $X = X(v,x)$},  obtained by
intersecting $dS_{d}$ with the hyperplanes $ X^{0} + X^{d} = e^
{v}$:
\begin{equation}
\left\{\begin{tabular}{lclcll}
 $X^{0} $&=& $\sinh  {v} + \frac 12 e^{ {v}} x^2 $  & & $ $ & $x^2 = {x^1}^2+ \cdots +{x^{d-1}}^2$ \cr
$X^{i} $&=& $e^{ {v}} x^i $ & & $
 $& $
 { i=1,...,d-1}$\cr
 $X^{d}$&=&$ \cosh  {v} - \frac 12 e^{ {v}} x^2$ &   &  &
\label{coordinates}
\end{tabular}\right. .
 \end{equation}
For real values of the parameters it only covers the part $\Pi$ of
the dS manifold which belongs to the half-space $\{X^0 + X^{d}>0
\}$ of the ambient space. Each slice $\Pi_v$ (or ``horosphere'')
is a (flat) paraboloid. The scalar product (1) and the dS metric
can then be rewritten as follows:
\begin{eqnarray}
&& X\cdot X' = -\cosh( {v}- {v} ')  + \frac 12 e^{ {v}+ {v}'}
\le(x-x'\ri)^2,
\label{7}\\
&& {\mathrm d}s^2 =  {\mathrm d} {v}^2 - e^{2 {v}} {\mathrm d}x^2
\label{metric1}
\end{eqnarray}
 Eq. (\ref{7}) implies that
\begin{equation}
(X(v,x)-X(v,x'))^2 = -e^{2v}(x-x')^2. \label{lll}
\end{equation}
This in turn implies that any slice $\Pi_v$ is conformal to a
Euclidean plane.

The de Sitter manifold has a boundary at timelike infinity. This
can be easily understood by using a Penrose diagram. Another
visualization can be obtained by taking the large $\tau$
asymptotics in eq. (\ref{sphericcoordinates}):
\begin{equation}
\left\{
\begin{tabular}{lclc}
$X^{0}$  &$\simeq$& $ \pm \, e^{|\tau|} $ & \cr
$X^{i} $  &$\simeq$& $\ \ e^{|\tau |}\ {\omega}^i $ & $i=1,...,d$,
\ \ $e^2 = {{\omega}^1}^2+ \cdots +{{\omega}^{d}}^2=1$
\end{tabular}\right.
\label{sphericcoordinatesas}
\end{equation}
It follows  that the ambient space  light-cone $$C = C_+\cup C_-=
{\cal C}_{1,d}= \{ \eta =(\eta^0,...,\eta^{(d+1)});\ {\eta^0}^2-
{\eta^1}^2- \cdots- {\eta^d}^2 =0 \}$$ can also be looked at as
the boundary at timelike infinity of the de Sitter manifold. We
will use the notation ${\cal C}_{1,d}$ to distinguish the
boundary where the asymptotic theory will live from the light-cone
itself, which has rather the interpretation of momentum space
\cite{Bros:1994dn,Bros:1996js}. The invariance group of the cone
${\cal C}_{1,d}$, which is also a copy of $SO_0(1,d)$, will be
interpreted as the euclidean conformal group
\cite{Luscher:1975ez}.

By adapting the covering parametrization
(\ref{sphericcoordinates}) of $ {dS}_{d}$ to the case of its
asymptotic cone ${\cal C}_{1,d}= \{ \eta
=(\eta^0,...,\eta^{(d+1)});\ {\eta^0}^2- {\eta^1}^2- \cdots-
{\eta^d}^2 =0 \}, $ one readily obtains the following
parametrization:
\begin{equation}
\left\{\begin{tabular}{lclcll} $\eta^{0} $  &=& $r  $ & \cr
$\eta^{i} $  &=& $r {\omega}^i $                   & ${
i=1,...,d}$ &
 \label{conecoordinates}
\end{tabular}\right. .
\end{equation}
with ${{\omega}^1}^2 + \ldots + {{\omega}^d}^2 = 1$ and $r \ge 0$,
or in brief: $\eta = \eta[r,{\omega}]$.

By taking the intersection of ${\cal C}_{1,d}$ with the family of
hyperplanes with equation $\eta^0 + \eta^{d} = e^v$, one obtains
the analogue of the horocyclic parametrization
(\ref{coordinates}), namely:
\begin{equation}
\left\{\begin{tabular}{lclcll}
$\eta^{0}$&=&$\frac 12 e^v (1- x^2) $\cr
 $\eta^{i} $&=& $e^{ {v}} x^i  $ & ${ i=1,...,d-1}$\cr
 $\eta^{d} $&=& $ \frac 12 e^v (1+ x^2) $
 &\ \  $x^2 ={ x^0}^2- {x^1}^2- \cdots -{x^{d-1}}^2$
\label{horocoordinates}
\end{tabular}\right.
\end{equation}
which implies the following identity (similar to (\ref{7}))
between quadratic forms:
\begin{equation}
(\eta -\eta')^2 = -e^{v+v'} (x-x')^2 \label{horocausal}
\end{equation}

By taking Eqs. (\ref{conecoordinates}) into account, one then
sees that these formulae correspond to the embedding of Euclidean
space into the the cone ${\cal C}_{1,d}$ namely one has (in view
of the identification $ \eta^0 +\eta^{d} = e^v = r( {\omega}^d +
1)$):
\begin{equation}
 x^i = \frac{\eta^{i}}{\eta^0+ \eta^{d}}
= \frac {{\omega}^i} {{\omega}^d+1}. \label{embed}
\end{equation}

In section (\ref{oo}) we shall prove that boundary theories
obtained as a certain limit from theories living in the bulk, are
$SO_0(1,d)$ symmetric (i.e. have the euclidean conformal
invariance). In this sense, boundary theories have the same
symmetry of bulk de Sitter theories.

\section{dS Quantum Field Theory}

One of the possible formulations of the AdS/CFT correspondence
states a duality between a perturbative (tree level) theory on
AdS and a non perturbative CFT on the boundary
\cite{Maldacena:1997re}. Nothing similar can at the moment be said
for the de Sitter case and therefore we must consider general
(non-perturbative) de Sitter theories.

A general approach to those theories has been developed in recent
years based on very general principles
\cite{Bros:1996js,Bros:1998ik} and we give here a very short
account of what is needed for the present purpose.  Various
consequences of these general principles have been derived in
\cite{Bros:1998ik} and most of the well-known properties of the
Wightman distributions in the Minkowskian case \cite{Streater},
including the PCT symmetry, hold without change in the de
Sitterian case under the assumptions specified below (see
\cite{Bros:1998ik} for a detailed account).

Before entering in the discussion, two important remarks are
however in order:
\begin{enumerate}
\item  the physical interpretation of the axioms specified below
is the thermal one. A geodesical observer will perceive the
``vacuum'' as populated by a thermal bath of particles, but it
has to be stressed that we are talking here of an interacting
quantum field theory;

\item the Reeh-Schlieder property holds. This
property says that the application to the vacuum vector  of the
polynomial field algebra of any open set in the de Sitter
manifold yields a dense set of the Hilbert space of the theory.
This reduces to irrelevance all the argument based on the presence
of observer's horizons that are used to discard the regions of the
de Sitter universe that are not accessible classically.
\end{enumerate}

Let us consider therefore a general QFT on $ {dS}_{d}$; for
simplicity we limit the present discussion to one scalar field
$\Phi(X)$. According to the general reconstruction procedure
\cite{Streater}, a theory is completely determined by the set of
all $n$-point vacuum expectation values (or ``Wightman
functions'')  of  the field $\Phi$, given as distributions on the
corresponding product manifolds ${dS}_{d}^n$:
\begin{equation}
{\cal W}_{n}(X_1, \ldots X_n)  \label{npoint}
\end{equation}
An important class of fields, which can be explicitly constructed
in a Fock space, is the class of ``generalized free fields'';
these fields are completely determined by their two-point
function ${\cal W}_{2}(X_1, X_2)$. In particular, the
Klein-Gordon fields are those for which ${\cal W}_{2}(X_1,X_2)$
satisfies the corresponding field equation w.r.t. both points. Of
course there are in general infinitely many inequivalent
solutions to this problem (encoded in the choice of ${\cal W}_2$)
and one has to select the meaningful ones on the basis of some
physical principle.

Let us denote $\DD_n$  the space of functions on $dS_d^n$
infinitely differentiable and with compact support. As in the
Minkowskian case, the Borchers algebra $\BB$ is defined as the
tensor algebra over $\DD = \DD_1$. Its elements are terminating
sequences of test-functions $f = (f_0, f_1(X_1),\ldots,
f_n(X_1,\ldots,X_n),\ \ldots)$, where $f_0 \in {\mathbb C}$ and
$f_n \in \DD_n$ for all $n \ge 1$, the product and $\star$
operations being given by $$ (fg)_n = \sum_{p,\ q \in \bN \atop
p+q = n}\ f_p \otimes g_q,\ \ \ \ (f^\star)_n (X_1,\ldots,\ X_n)
= \ovl{f_n (X_n,\ldots,\ X_1)}. $$ The action of the de Sitter
group  on $\BB$ is defined by $ f \mapsto f_{\{\Lambda_r\}}$,
where
\begin{equation}
f_{\{\Lambda_r\}} = (f_0,
f_{1\{\Lambda_r\}},\ldots,f_{n\{\Lambda_r\}},\ldots),\ \ \ \
f_{n\{\Lambda_r\}} ({x_{1}},\ldots,x_{n}) =f_{n} ({\Lambda_r}
^{-1}{x_{1}},\ldots,{\Lambda_r}^{-1}x_{n}), \label{Lambdaact}
\end{equation}
$\Lambda_r$ denoting any (real) de Sitter transformation. \vskip
5pt

A quantum field theory is specified by a continuous linear
functional $\WW$ on $\BB$, i.e. a sequence $\{\WW_n \in
\DD'_n\}_{n \in \bN}$ where $\WW_0 =1$ and the $\{\WW_n\}_{n
>0}$ are distributions (Wightman functions) required to possess
the following properties:

\begin{enumerate}
\item (Covariance). Each ${\cal W}_{n}$ is de Sitter invariant, i.e.
\begin{equation}
\langle {\cal W}_{n},\ f_{n\{\Lambda_r\}}\rangle = \langle {\cal
W}_{n},\ f_{n} \rangle \label{cov}
\end{equation}
for all de Sitter transformations $\Lambda_r$. (This is
equivalent to saying that the functional $\cal W$ itself is
invariant, i.e. $ {\cal W}(f) = {\cal W}(f_{\{\Lambda_r\}})$ for
all $\Lambda_r$).

\item (Locality)
\begin{equation}
{\cal W}_{n}({X_{1}},\ldots,X_{j},X_{j+1},\ldots,X_{n}) ={\cal
W}_{n}({X_{1}},\ldots,X_{j+1},X_{j},\ldots,X_{n})
\end{equation}
if $(X_{j}-X_{j+1})^{2}<0$.
\item (Positive Definiteness). For each $f \in \BB$,
$\WW(f^\star f) \ge 0$. Explicitly, given $f_{0} \in {{\mathbb
C}}, f_{1} \in \DD_1,\ldots,$ $f_{k} \in \DD_k,$ then
\begin{equation}
\sum_{n,m=0}^{k}\langle {\cal W}_{n+m},\ f_{n}^\star\otimes
f_{m}\rangle\geq 0. \label{posit}
\end{equation}
\end{enumerate}
The latter property should be possibly relaxed to treat de Sitter
gauge QFT.

As in the Minkowskian case the GNS construction yields a Hilbert
space $\HH$, a unitary representation $\Lambda_r \mapsto
U(\Lambda_r)$ of $SO_0(1,d)$, a vacuum vector $\Omega \in \HH$
invariant under $U$, and an operator valued distribution $\phi$
such that
\begin{equation}
\WW_n(X_1,\ldots,\ X_n) = (\Omega,\
\phi(X_1)\ldots\phi(X_n)\,\Omega).
\end{equation}
The GNS construction also provides the vector valued distributions
$\Phi_n^{(b)}$ such that
\begin{equation}
\langle \Phi_n^{(b)},\ f_n \rangle = \int f_n(X_1,\ldots,\
X_n)\,\phi(X_1)\ldots\phi(X_n)\,\Omega\, d\sigma(X_1)\ldots
d\sigma(X_n) \label{vecvaldis}
\end{equation}
and a representation ${  f} \to \Rep({f})$ (by unbounded
operators) of ${\cal B} $ of which the field $\phi$ is a special
case: $ \phi({ f_1})= \int \phi(X)f_1(X) d\sigma(X) =
\Rep\left((0,f_1,0,\ldots)\right)$. For every open set ${\cal O}$
of $dS_d$ the corresponding polynomial algebra ${\cal P}({\cal
O})$ of the field $\phi$ is then defined as the subalgebra of
$\Rep({\cal B})$ whose elements $\Rep(f_0,f_1,\ldots,f_n,\ldots)$
are such that for all $n\geq 1\;$ supp$f_n(x_1,\ldots,
x_n)\subset {\cal O}^{n}$. The set ${\rm D} = {\cal
P}(dS_d)\Omega$ is a dense subset of ${\cal H}$ and one has (for
all elements $\Rep({f}), \Rep({g}) \in {\cal P}(dS_d)$):
\begin{equation}
\WW(f^\star g) = (\Rep({f})\Omega,\ \Rep({g})\Omega).
\label{borch}
\end{equation}

The properties 1-3 are literally carried over from the
Minkowskian case, but no literal or unique adaptation exists for
the usual spectral property. In the $(d+1)$-dimensional
Minkowskian case, the latter is equivalent to the following: for
each $n \ge 2$, $\WW_n$ is the boundary value in the sense of
distributions of a function holomorphic in the tube
\begin{equation}
{\rm T}_n = \{Z= (Z_1,\ldots,\ Z_n) \in \bC^{n(d+1)}\ :\ \Im
(Z_{j+1} - Z_j) \in V_+ ,\;1\leq j\leq n-1 \} . \label{tubular}
\end{equation}
In the case of the de Sitter space $dS_d$ (embedded in ${{\mathbb
R}}^{d+1}$), a natural substitute for this property is to assume
that $\WW_n$ is the boundary value in the sense of distributions
of a function holomorphic in
\begin{equation}
\TT_n = dS^{(c)n}_d \cap {\rm T}_n. \label{tuboidal}
\end{equation}
It has been shown  that $\TT_n$ is a domain and a tuboid
\cite{Bros:1996js,Bros:1998ik}, namely a domain which is bordered
by the reals in such a way that the notion of ``distribution
boundary value of a holomorphic function from this domain''
remains meaningful. It is thus possible to impose:
\begin{enumerate}
\setcounter{enumi}{3}
\item (Weak spectral condition).
For each $n > 1$, the distribution ${\cal W}_{n}$ is the boundary
value of a function ${\rm W}_{n}$ holomorphic in the subdomain
$\TT_n$ of $dS^{(c)n}_d$.
\end{enumerate}

\section{Correspondence with  conformal field theories on ${\cal
C}_{1,d}$: dimensional boundary conditions at infinity.}
\label{oo}

In order to obtain correlation functions on the boundary of dS
spacetime we are led to postulate a certain type of behavior at
infinity for the Wightman functions which we call ``dimensional
boundary conditions at infinity''.

By making use of the coordinates (\ref{sphericcoordinates}) we say
that a QFT on $ {dS}_{d}$ is of asymptotic (complex) dimension
$\Delta$ if the following limits exist in some sense:
\begin{enumerate}
\setcounter{enumi}{4} \item{(Dimensional boundary conditions at
infinity)}
\begin{eqnarray}
 & \lim_{\tau \to {+\infty}} &|\sinh {\tau}|^{n\Delta}
\ {\cal W}_n (X_1[\tau,{\omega}_1],...,X_n[\tau,{\omega}_n]) \cr
&&= {\cal W}_n^{\infty}({\omega}_1,...,{\omega}_n) \label{15}
 \end{eqnarray}
\end{enumerate}
In words: we take the restriction of the $n$-point function to
the manifold $S(\tau)^n$ (i.e. set all the times $\tau_j=\tau$),
multiply by the indicated factor and take the limit to infinity.

The first thing to be shown is  that the above condition is
meaningful, since it is not true in general that a distribution
${\cal W}_{n}(X_1, \ldots X_n)$ can be restricted to a submanifold
of ${dS}_{d}^n$.

Our spectral condition 4. implies that this can be done at equal
times  for  noncoinciding points. The argument is based on the
existence of an analytic continuation of the Wightman $n$-point
functions to corresponding primitive domains of analyticity.

Indeed for each permutation $\pi$ of $(1,\ldots,n)$, the permuted
Wightman distribution
\begin{equation}
{\cal W}^{(\pi)}_{n}(X_{1},\ldots,X_{n})= {\cal
W}_{n}(X_{\pi(1)},\ldots,X_{\pi(n)})
\end{equation}
is the boundary value of a function ${\rm
W}^{(\pi)}_{n}(Z_{1},\ldots,Z_{n})$ holomorphic in the "permuted
tuboid"
\begin{eqnarray}
&{\cal T}^{\pi}_n=& \{{ Z} = ({ Z}_{1},\ldots,{ Z}_{n});\;{
Z}_{k}= { X}_{k} + i{ Y}_{k}\in dS^{(c)}_d,\;1\leq k \leq n; \cr
&& \;{ Y}_{\pi(j+1)}- { Y}_{\pi(j)}\in V^{+},\;1\leq j \leq n-1\}
\label{tubular11}
\end{eqnarray}
If two permutations $\pi$ and $\sigma$ differ only by the exchange
of the indices $j$ and $k$, then $\WW_\pi$ and $\WW_\sigma$
coincide in
\begin{equation}
\RR_{j k} = dS_d^n \cap {\rm R}_{jk},\ \ \ {\rm R}_{j k} = \{X \in
\bR^{n(d+1)}\ :\ (X_j-X_k)^2 < 0\}.
\end{equation}
In particular all the permuted Wightman distributions coincide in
the intersection $\Omega_n$ of all the $\RR_{j k}$, and it
follows that they all are boundary values of a common function
$\makebox{\Eul W}_{n}(Z_{1},\ldots Z_{n})$, holomorphic in a {\it
primitive analyticity domain} ${\cal D}_{n}$. $\makebox{\Eul
W}_{n}$ is the common analytic continuation of all the
holomorphic functions ${\rm W}^{(\pi)}_{n}$ and the domain ${\cal
D}_{n}$ is the union of all the permuted tuboids ${\cal
T}^{\pi}_{n}$ and of the local tuboids associated (by the
edge-of-the-wedge theorem) with finite intersections of the
$\RR_{j k}$. It is self-evident that any n-tuple $
\{X_1[\tau,\omega_1],...X_n[\tau,\omega_n]\}$ such that $X_i \not
= X_j$ for any $i \not = j$ belongs to such  primitive domain of
analyticity and  actually it belongs to $\Omega_n$; therefore the
previous restriction can be considered.

\vskip 0.4cm Let us now consider a general QFT on $ {dS}_{d}$
whose Wightman functions ${\cal W}_n$ satisfy dS invariance
together with the properties  described in the previous section,
with the possible exception of the positive-definiteness property
3. In view of the asymptotics we can construct the following set
of $n-$point distributions $ {\cal E}_n (\eta_1,...,\eta_n)$ on $
{\cal C}_{1,d}^+$ \cite{Luscher:1975ez,Bertola:1999sd}:
\begin{equation}
 {\cal E}_n (\eta_1,...,\eta_n) =
 (r_1 \cdots r_n)^{-\Delta}
{\cal W}_n^{\infty}({\omega}_1,...,{\omega}_n). \label{19}
\end{equation}

\vskip 0.4cm We are now going to establish that the
$SO_0(1,d)$-invariance of the de Sitter $n-$point functions,
together with the asymptotic boundary condition imply the
$SO_0(1,d)$-invariance invariance of the correlation functions $
{\cal E}_n $ now interpreted as euclidean conformal
transformations of ${\cal C}_{1,d}$:
\begin{equation}
 {\cal E}_n (g\eta_1,...,g\eta_n) =
 {\cal E}_n (\eta_1,...,\eta_n)
\label{qmcf}
\end{equation}
for all $g$ in $SO_0(1,d)$. A part of this invariance is trivial
in view of the limiting procedure: it is the invariance under  the
spatial orthogonal group $SO(d)$ leaving $\eta^0$ unchanged.

In order to show that the invariance condition (\ref{qmcf}) holds
for all $g$ in $G$, it remains to show that it holds for all
one-parameter subgroups of pseudo-rotations in the $(0,i)-$planes
with $i=1,...,d$. Closely following the steps indicated in
\cite{Bertola:1999sd} let us consider the case with e.g. $i=1$
and associate with the corresponding subgroup $G_{0,1}$ of
pseudo-rotations the following parametrization
$X=X\{\rho,\psi,u\}$ (with $u= (u^2,...,u^{d})$) of $ {dS}_{d}$ :
\begin{eqnarray}
&&\left\{\begin{tabular}{lclcll} $X^{0} $  &=& $\rho \cosh \psi $
& &$\rho>0$ \cr
$X^{1} $  &=& $\rho \sinh \psi $ & \cr
$X^{i}$ &=&$ \sqrt {{\rho}^2 - 1} \   u^i$& & ${ i=2,...,d}$ &
${u^2}^2 +\cdots +{u^d}^2 =1$ \label{pseudodS}
\end{tabular}\right.  .
\end{eqnarray}

Correspondingly we have the following parametrization   $\eta
=\eta\{\rho,\psi,u\}$  for the cone $ {\cal C}_{1,d}^+$:
\begin{eqnarray}
&&\left\{\begin{tabular}{lclcll} $\eta^{0} $  &=& $\sigma \cosh
\psi $ & & $\sigma>0$ \cr
$\eta^{1} $  &=& $\sigma \sinh \psi $ & \cr
$\eta^{i}$ &=&$ \sigma \   u^i$& & ${ i=2,...,d}$ & $ {u^2}^2
+\cdots +{u^d}^2 =1$ \label{pseudocone}
\end{tabular}\right.
\end{eqnarray}
For $g \in G_{0,1},$ the invariance condition (\ref{qmcf}) to be
proven can be written as follows (with the simplified notation $
{\cal E}_n (\eta_1,...,\eta_n) =  {\cal E}_n (\eta_j)$):
\begin{equation}
 {\cal E}_n (\eta_j\{\sigma_j,\psi_j +a,w_j\})
=  {\cal E}_n (\eta_j\{\sigma_j,\psi_j,w_j\}) \label{psi-invar}
\end{equation}
for all real $a$. Now in view of the definition (\ref{19}) of $
{\cal W}_n (\eta_j)$ and of the relations between the sets of
parameters $(r,\tau,{\omega})$ and $(\rho,\psi,u)$ obtained by
identification of the expressions (\ref{conecoordinates}) and
(\ref{pseudocone}) of $\eta$, the invariance condition
(\ref{psi-invar}) to be proven is equivalent to the following
condition for the asymptotic forms of the dS $n-$point functions
${\cal W}_n^{\infty}$ (for all $a$):
\begin{eqnarray}
&& \prod_{1 \leq k \leq n} \left( \sigma_k\cosh \psi_k
\right)^{-{\Delta}}
{\cal W}_n^{\infty} \left(\left[{\tanh \psi_j}, \frac {u_j}{\cosh
\psi_j}
 \right]\right)= \cr
&& \prod_{1 \leq k \leq n}
 \left(\sigma_k (\cosh(\psi_k+a))
\right)^{-{\Delta}}{\cal W}_n^{\infty} \left(\left[{\tanh (\psi_j
+ a)}, \frac {u_j}{\cosh (\psi_j +a)}
 \right]\right). \label{30}
\end{eqnarray}
Comparing the parametrizations (\ref{sphericcoordinates}) and
(\ref{pseudodS}) of $ {dS}_{d}$ we obtain the following relations:
\begin{eqnarray}
&& \sinh \tau = \rho \cosh \psi, \ \ \
\omega^1 =\frac{\rho \sinh \psi}{\sqrt{1+\rho^2 \cosh^2 \psi}}, \
 \ \
\omega^i =\frac{\rho \sinh \psi u^i}{\sqrt{1+\rho^2 \cosh^2 \psi}}
\label{31}
\end{eqnarray}
This implies that it is equivalent to take the limits in Eq.
(\ref{15}) for $\rho_j$ (instead of $\tau$) tending to infinity
and at fixed value of $\psi_j$ and $u_j$, after plugging  the
expressions (\ref{31}) of $\tau = \tau_j$ and ${\omega}_j$ into
both sides of Eq. (\ref{15}):

\begin{eqnarray}
&& \lim_{\tau \to {\infty}} \left|(\rho_1\cdots \rho_n)^{\Delta}
{\cal W}_n (X_j\{\rho_j,\psi_j,u_j\})\right. -\cr &&
\!\!\!\!\!\left. \prod_{1 \leq k \leq n} (\cosh\psi_k)^{-\Delta}
{\cal W}_n^{\infty}
\left(\left[ \tanh{\psi_j},
\frac{\rho \sinh \psi_j}{\sqrt{1+\rho^2 \cosh^2 \psi_j}},
\frac{\rho \sinh \psi u^i_j}{\sqrt{1+\rho^2 \cosh^2 \psi }}
\right]\right)\right|  = 0. \label{34}
\end{eqnarray}
If we now also consider the vanishing limit of the same
difference after the transformation $\psi_j \to \psi_j +a$ has
been applied, and take into account the fact that, by assumption,
the first term of this difference has remained unchanged, we
obtain the following relation:
\begin{equation}
 \begin{array}l \lim_{\tau \to {+\infty}}
\left| \prod_{1 \leq k \leq n} (\cosh\psi_k)^{-\Delta} {\cal
W}_n^{\infty}
\left(\left[ \tanh{\psi_j},
\frac{\rho \sinh \psi_j}{\sqrt{1+\rho^2 \cosh^2 \psi_j}},
\frac{\rho \sinh \psi u^i_j}{\sqrt{1+\rho^2 \cosh^2 \psi }}
\right]\right)\right. - \\
 \left. \prod_{1 \leq k \leq n}
\cosh(\psi_k+a)^{-\Delta} \ {\cal W}_n^{\infty}
\left(\left[ \tanh (\psi_j +a),
\frac{\rho \sinh(\psi_j + a)}
{\sqrt{1+\rho^2 \cosh^2 (\psi_j+a)}},
\frac{\rho \sinh (\psi+a) u^i_j}{\sqrt{1+\rho^2 \cosh^2 (\psi
+a)}} \right]\right)\right|
 = 0.
\end{array}
\end{equation}
In the latter, the limit can be taken separately in each term and
the resulting equality yields precisely the required
covariance relation (\ref{30}). \vskip10pt
 We stress again  that,
thanks to our general setting and in particular the spectral
condition 4., all the functions involved are of class ${\cal
C}^{\infty}$ with respect to all the variables
$(\rho_j,\psi_j,u_j)$ and all the limits are taken in the sense
of regular functions.

It would not have been possible to take restriction for points in
general position; in particular to send for instance one point to
minus infinity and the remaining points to plus infinity is not
allowed  since after a certain time all the points will enter the
future cone of the point moving to minus infinity and the
restriction would become meaningless.

The   procedure we have described (expressed by Eqs. (\ref{15})
and (\ref{19})) displays a general correspondence
\begin{equation}{\cal W}_n(X_1,\ldots,X_n)
\rightarrow {\mathcal E}_n(\eta_1,\ldots,\eta_n)\ .
\label{correspondence}
\end{equation}
The degree of homogeneity (dimension) $\Delta$ of ${\mathcal
E}(\eta_1,\ldots\eta_n)$ is equal to the asymptotic dimension of
the dS field $\Phi(X)$. The correspondence (\ref{correspondence})
can be completed by constructing $n-$point functions ${ E}_n$ on
the Euclidean space ${\mathbb E}^{d-1}$, expressed in terms  of
the ${\mathcal E}_n$ by the following formulae
\cite{Luscher:1975ez,Todorov}:
\begin{equation}
{E}_n(x_1,...,x_n)  = \Pi_{1 \leq j \leq n} (\eta_j^0 +
\eta_j^{d})^{\Delta}\   {\cal E}_n (\eta_1,...,\eta_n) .
\label{eucl}
\end{equation}
In the latter, the Euclidean variables $x_j$ are expressed in
terms of the cone variables $\eta_j$ as in Eq. ({\ref{embed}).
\section{Discussion}
Could we say that there exists a CFT associated to the
so-constructed euclidean conformal correlation functions? If the
original dS theory does not satisfy the positive-definiteness
property the best that one can do in general is to use the GNS
construction to build a linear space endowed with a (non-positive
or indefinite metric) inner product and an operator valued
distribution ${\cal O}(\eta)$ having such euclidean correlation
functions.

But even if one has positive-definiteness in the dS theory, i.e.
if the dS theory has a direct physical interpretation, the
construction does not guarantee that also the limiting euclidean
theory is positive definite and the GNS construction  gives an
Hilbert space, because coinciding points are not in general under
control (this is the unitarity property mentioned in \cite{stro}).

However this still does not settle the question of the physical
meaning of the asymptotic theory. Indeed it is well known that
the only meaningful notion of positivity for euclidean theories
to admit a direct physical interpretation is the so-called
Osterwalder-Schrader positivity or reflection positivity (and {\em
not} positive-definiteness); 
it is the only condition which allows the reconstruction
of quantities at real time through an appropriate 
Wick-rotation from the Euclidean 
$n$-point functions.     

Unfortunately there is no way to show that the asymptotic theory
has such a property, and actually in general it will 
not\footnote{This is in contrast with what would happen with 
Euclidean sections of a Minkowskian theory, if a similar construction
to the present one was performed. In that case reflection positivity
would hold. The reason for this difference
resides in the lack of translation invariance of the curved case.}, even if
in some lucky case this property may still hold. This generic
situation will be illustrated by the following free field examples
\cite{stro,witten}.

\section{Klein-Gordon fields}
Let us consider  now the de Sitter Klein-Gordon equation
\begin{equation}
\Box \phi +{m^{2} } \phi=0, \label{kgdS}
\end{equation}
where $\Box $ is the Laplace-Beltrami operator relative to the de
Sitter metric and $m^2$ is a complex number. It is possible to
solve in a coordinate-independent way \cite{Bros:1996js} by using
the previous embedding of the de Sitter hyperboloid in the
Minkowski ambient space. First of all one introduces plane  waves
solving the KG equation; these waves are similar to the
Minkowskian exponentials but with the important difference that
they are singular on $(d-1)$-dimensional light-like submanifolds
of $dS_d$. The physically relevant global waves can be defined as
analytic functions for $z$ in the tubular domains $ {\cal T}_{+}$
or $ {\cal T}_{-}$ of $dS^{(c)}_{d}$; for $Z\in {\cal T}_{+}$ or
$ Z \in {\cal T}_{-}$ we define
\begin{equation}
{\psi}_{i\nu}^{(d)} (Z ,\xi )= \left( {Z  \cdot {\xi}}
\right)^{-\frac{d-1}{2} + i\nu }, \label{dSwaves}
\end{equation}
where $\nu $ is a complex number and ${\xi} = (\xi^{0},\ldots
\xi^{d})$ belongs to $C_+$. The phase is chosen to be zero when
the argument is real and positive. Physical values of the
parameter $\nu$ are real (principal series of representations) or
purely imaginary with $|\nu| \leq \frac{d-1}{2}$ (complementary
series of represenatations), corresponding to a real and positive
$m^2$:
\begin{equation}
m^2 = \left(\frac{d-1}{2}\right)^2 + \nu^2 \, > \, 0.
\end{equation}
but we will study the limit for generic complex $\nu$.
The corresponding QFT is completely encoded in the two-point
function ${W}_\nu(X,X')$ which should be a distribution on
$dS_d\times dS_d$ satisfying the  conditions of locality de
Sitter invariance; {\em positive-definiteness will hold only for
physical values of $m^2$.}
${W}(X,X')$ should solve the KG  w.r.t. both variables:
\begin{equation}
\left(\Box_{X}  +{m^{2} }\right){ W}(X,X') =0,\qquad
\left(\Box_{X'}  +{m^{2} }\right){ W}(X,X') =0.
\end{equation}
There are infinitely many inequivalent  solutions to this problem,
but there is one preferred theory (for each value of the mass
$m$) which is usally referred to as the  ``Euclidean'' or
Bunch-Davies vacuum \cite{[BuD],Gibbons:1977mu,Bros:1996js}; what
is perhaps not so well known is that these fields can be directly
constructed in a manifestly de Sitter invariant way
\cite{Bros:1996js} by exploiting the previous dS plane waves.
Indeed it is possible to give a spectral analysis of the
two-point functions very similar to the Fourier analysis usually
done in the flat Minkowski case. This is constructed as follows:
for $Z\in {\cal T}_{-}$, $Z'\in {\cal T}_{+}$ the Wightman
function can be represented as a superposition of plane waves in
the complex domain ${\cal T}_{-}\times{\cal T}_{+}$
\cite{Bros:1996js}:
\begin{equation}
{W^ {d}_\nu}(Z,Z') = c_{d,\nu}\int_\gamma { {\psi}_{i\nu}^{(d)}
(Z,\xi) {\psi}_{-i\nu}^{(d)} (Z',\xi) d\mu_\gamma(\xi)}
\label{tpf}
\end{equation}
with
\begin{equation}
c_{d,\nu}=\,\frac{1}{2(2\pi)^{ d}}\,{ {{\Gamma\left(\frac{d-1}{2}+
i\nu\right)}{\Gamma\left({\frac{d-1}{2}}-i\nu\right)}}e^{-\pi\nu}}.
\end{equation}
The integration can be performed along any basis submanifold
$\gamma$ of the cone $C_+$ (i.e. a submanifold intersecting almost
all the generatrices of the cone) w.r.t. a corresponding measure
$d\mu_\gamma$ induced by the invariant measure on the cone. For
instance, one can integrate  on the  manifold
$\gamma_{d}=\gamma^+_{d}\cup\gamma^-_{d} =\{ \xi \in C_+: \xi^{d}
= 1 \}\cup \{ \xi \in C_+ : \xi^{d} = -1 \}$, which is a pair of
hyperboloids; in this case the measure $d\mu_\gamma$ looks like
the Lorentz invariant measure on the mass shell. For the
spherical basis $\gamma_{0}= \{ \xi \in C^+: \xi^{0} = 1\}$ the
measure $d\mu_\gamma$ is exactly  the rotation invariant measure
(on the sphere).

The function ${W^ {d}_\nu}$ manifestly solves the (complex) de
Sitter Klein-Gordon equation in both variables, and is analytic
in the domain ${\cal T}_{-}\times{\cal T}_{+}$. It can be shown
that it is actually a function of the de Sitter invariant
variable $(Z-Z')^{2} = -2 - 2Z\cdot Z'$. This property allows the
explicit computation
\begin{equation}
{W^ {d}_\nu}(Z,Z')
 =
 \label{wig1}
\frac{1} {2(2\pi)^{\frac{d}{2}}}\Gamma\left(\frac{d-1}{2}
+i\nu\right)
            \Gamma\left(\frac{d-1}{2} -i \nu\right)
 \,
\,((Z\cdot Z')^2-1)^{-\frac{d-2}{4}}\,
P^{-\frac{d-2}{2}}_{-\frac{1}{2} + i\nu}\left({Z\cdot Z'}\right),
\label{Pdef}
\end{equation}
where $P^{-\frac{d-2}{2}}_{-\frac{1}{2} + i\nu}(\zeta)$ is
Legendre function of the first kind \cite{Bateman}. At vanishingly
short distances the Wightman function has the local Hadamard
universal behaviour:
\begin{equation}
{W^ {d}_\nu}(Z,Z') \simeq
\frac{\Gamma(\frac{d-2}{2})}{2(2\pi)^{\frac{d}{2}}}[-(Z-Z')^2]^{-\frac{d-2}{2}}.
\label{Mink}
\end{equation}
By equation (\ref{wig1}) one sees that ${W^ {d}_\nu} (z,{{z'}})$
is maximally analytic, i.e. can be analytically continued in the
``cut-domain'' $dS^{(c)}_d\times dS^{(c)}_d \setminus$
$\{(z,{{z'}}) \in dS^{(c)}_{d}\times dS^{(c)}_d: (Z-Z')^2  \geq
0\}$. Furthermore, ${{W^ {d}_\nu}}(Z,Z')$ satisfies in this
cut-domain the complex covariance condition: $ {{W^ {d}_\nu}}
(gZ,gZ')= {{W^ {d}_\nu}} (Z,Z')$ for all $g$ in the complex de
Sitter group\footnote{These properties are not restricted to
Klein-Gordon fields and are actually true for any two-point
Wightman function $W$ satisfying our spectral condition
\cite{Bros:1996js}}.

As a function of the parameter $\nu$ the two-point function
$W_\nu$ is analytic  and symmetric: $W^d_{-\nu} = W^d_{\nu}$.

\subsection{Boundary theories from KG fields with a complex mass}
For large values of the argument $W^d_{\nu}(\zeta)$  has the
following asymptotics \cite{Bateman} :

\begin{equation}
{ W}_{\nu}(\zeta)  \sim \frac{2^{-i\nu }\Gamma\left(\frac{d-1}{2}
+i\nu\right)\Gamma(-i\nu)} { 2(2\pi)^{\frac{d+1}{2}}}\ \ \zeta^{
-\frac{d-1}{2}-i\nu} \label{Pinf>}  \ \ \ \ \ \ \ \makebox{for
Im} \ \nu > 0
\end{equation}
\begin{equation} W_{ \nu}(\zeta)  \sim
\frac{2^{i\nu }\Gamma\left(\frac{d-1}{2}
-i\nu\right)\Gamma(i\nu)} { 2(2\pi)^{\frac{d+1}{2}}}\ \   \zeta^{
-\frac{d-1}{2}+i\nu} \label{Pinf<} \ \ \ \ \ \  \ \  \makebox{for
Im} \ \nu < 0
\end{equation}
In the relevant case when  $\Im \nu = 0$, that corresponds to
physical KG fields of the principal series, the two terms are of
the same order and both contribute.

When Im $\nu > 0$ (resp. Im $\nu < 0$) the two-point function and
thereby all the $n-$point functions of the corresponding
Klein-Gordon field satisfy the dimensional boundary conditions at
infinity with dimension $\Delta = \frac{d-1}{2} + i\nu$ (resp.
$\Delta = \frac{d-1}{2} - i\nu$). Indeed, $\makebox{for Im} \,\nu
> 0$
\begin{eqnarray}
{W}_\nu^{\infty}({\omega},{\omega}') & = &\lim_{\tau \to \infty}
({\sinh^2 \tau})^{\frac{d-1}{2} + i
\nu}W_{\nu}(Z[\tau,{\omega}],Z'[\tau,{\omega}']) =\cr &=&
\frac{\Gamma\left(\frac{d-1}{2} +i\nu\right)\Gamma(-i\nu)} {
2(2\pi)^{\frac{d+1}{2}}}\ \ 2^{-i\nu }  (1-\omega\cdot
\omega')^{-i\nu -\frac{d-1}{2}}.
  \label{limit1}
\end{eqnarray}
The two-point function of the  conformal field ${ \mathcal
O}(\eta)$ on the cone  corresponding to (\ref{limit1}) is then
constructed by following the prescription of Eq.(\ref{19}), which
yields
\begin{equation}
   { \cal E}_{\nu}  (\eta ,\eta') =
  (r r')^{-\frac{d-1}{2} - i \nu}
{ W}_{\nu}^{\infty}({\omega},{\omega}') =
\frac{\Gamma\left(\frac{d-1}{2} +i\nu\right)\Gamma(-i\nu)} { 4
\pi^{\frac{d+1}{2}}}\ \  \left[-(\eta - \eta')^2\right]^{-i\nu
-\frac{d-1}{2}} \label{limit2}
\end{equation}
Correspondingly, we can deduce from (\ref{limit2}) the expression
of the two-point function of the associated euclidean two-point
function on $\mathbb E^{d-1}$; by taking Eqs. (\ref{eucl}) and
(\ref{horocausal}) into account, we obtain:
\begin{equation}
 {\rm E}_{\nu}^{d-1}(x,x') =  e^{(v+v')
(-\frac{d}{2} -i\nu)} { W}_{\nu} \left( \eta\left(v,x \right) ,
\eta'\left(v',x'\right)\right)= \frac{\Gamma\left(\frac{d-1}{2}
+i\nu\right)\Gamma(-i\nu)} { 4 \pi^{\frac{d+1}{2}}}\ \ \left[(x-
x')^2\right]^{-i\nu -\frac{d-1}{2}}  \label{gurru}
\end{equation}
Similar results hold for  $\makebox{for Im} \,\nu < 0$.

\subsection{Physical case: the complementary series}
In this case one has $\nu = i \lambda$ with $0<\lambda <
\frac{d-1}{2}$.

\begin{equation}
 {\rm E}^{d-1}_{\lambda}(x,x') = \frac{\Gamma\left(\frac{d-1}{2}
- \lambda\right)\Gamma(\lambda)} { 4 \pi^{\frac{d+1}{2}}}\ \
\left[(x- x')^2\right]^{\lambda -\frac{d-1}{2}}  \label{gurru1}
\end{equation}
This two-point function does satisfy positive definiteness
exactly when $\lambda $ satisfies the above condition and we can
construct an Hilbert space out of it in the usual GNS way
\cite{stro}.

One also checks easily that OS positivity holds when $\lambda
\leq 1$. It is interesting to note that the two bounds coincide in
the three dimensional case where the boundary theory has the full
infinite-dimensional conformal invariance. In the two-dimensional
case theories having the OS positivity arise from non-unitary de
Sitter theories. It might also be possible to get  other boundary
CFT's violating the bound $\lambda \leq 1$. These theories would
arise however as limit of exotic de Sitter theories, which do not
satisfy  de Sitter locality.
\subsection{Physical case: the principal series}

These are  dS Klein-Gordon theories  corresponding to real values
of $\nu$ and are the theories which have a standard flat limit
\cite{Bros:1996js} (while theories of the complementary series
disappear in that limit). Unfortunately they do not satisfy our
asymptotic dimension property. The best one can do is to give a
small imaginary part to $\nu$ and then apply the previous
construction. One sees that a theory of the complementary series
can be associated this way to two boundary theories which however
have complex dimensions and satisfy neither positive definiteness
nor OS positivity.
\vskip10pt
 \noindent{\bf Acknowledgments:} U.M. thanks Vincent
Pasquier for several useful discussions.

\end{document}